\newcommand{\lya}{\mbox{Ly$\alpha$}}

\newcommand{\hgamma}{\mbox{H$\gamma$}}

\newcommand{\hI}{\mbox{H~{\sc i}}}
\newcommand{\hII}{\mbox{H~{\sc ii}}}

\newcommand{\oII}{\mbox{O~{\sc ii}}}
\newcommand{\oIII}{\mbox{O~{\sc iii}}}

\newcommand{\cII}{\mbox{C~{\sc ii}}}
\newcommand{\cIII}{\mbox{C~{\sc iii}}}

\newcommand{\kms}{\mbox{km\,s$^{-1}$}}

\newcommand{\msun}{\mbox{M$_\odot$}}

\newcommand{\fesclyc}{\mbox{$f_\mathrm{esc}^\mathrm{LyC}$}}

\newcommand{\Lstar}{\mbox{$L^\star$}}

\newcommand{\nh}{\mbox{$n_\mathrm{H}$}}

\newcommand{\zsys}{\mbox{$z_\mathrm{sys}$}}

\newcommand{\xiion}{\mbox{$\xi_{\mathrm{ion}}$}}

\newcommand{\xHI}{\mbox{$\bar{x}_\mathrm{HI}$}}

\newcommand{\vigm}{\mbox{$\Delta v_\mathrm{IGM}$}}
\newcommand{\dvred}{\mbox{$\Delta v_\mathrm{red}(\mathrm{Ly}\alpha)$}}

\newcommand{\Rbub}{\mbox{$R_\mathrm{B}$}}
\newcommand{\Rstrom}{\mbox{$R_\mathrm{S}$}}

\documentclass[a4paper,fleqn,usenatbib,twocolumn]{aastex631}
\usepackage{newtxtext,newtxmath}
\usepackage[T1]{fontenc}
\usepackage{xcolor}
\usepackage{longtable}
\usepackage{tablefootnote}
\usepackage{graphicx}	
\usepackage{amsmath}	

\received{March 10, 2023}
\revised{XXX}
\accepted{\today}
\submitjournal{ApJ}

\shorttitle{Lyman alpha and IGM bubbles at high-z}
\shortauthors{M. J. Hayes et al.}

\graphicspath{{figs/}}

\begin{document}

\title{ON THE SIZES OF IONIZED BUBBLES AROUND GALAXIES DURING THE REIONIZATION
EPOCH.\\  THE SPECTRAL SHAPES OF THE LYMAN-ALPHA EMISSION FROM GALAXIES.}

\correspondingauthor{Matthew J. Hayes}
\email{matthew@astro.su.se}

\author[0000-0001-8587-218X]{Matthew J. Hayes}
\affiliation{Stockholm University, Department of Astronomy and Oskar Klein
Centre for Cosmoparticle Physics, AlbaNova University Centre, SE-10691,
Stockholm, Sweden.}
\author[0000-0002-9136-8876]{Claudia Scarlata}
\affiliation{Minnesota Institute for Astrophysics, School of Physics and
Astronomy, University of Minnesota, 316 Church Str. SE, Minneapolis,MN 55455,
USA}

\begin{abstract} 
We develop a new method to determine the distance between a high-redshift
galaxy and a foreground screen of atomic hydrogen.  In a partially neutral
universe, and assuming spherical symmetry, this equates to the radius of a
ionized `bubble' (\Rbub) surrounding the galaxy.  The method requires an
observed \lya\ equivalent width, its velocity offset from systemic, and an
input \lya\ profile for which we adopt scaled versions of the profiles observed
in low-$z$ galaxies. We demonstrate the technique in a sample of 23 galaxies at
$z>6$, including eight at $z=7.2-10.6$ recently observed with JWST.  Our model
estimates the emergent \lya\ properties, and the foreground distance to the
absorbing IGM.  We find that galaxies at $z>7.5$ occupy smaller bubbles ($\sim
0.5-1$~pMpc) than those at lower-$z$.  With a relationship that is secure at
99\% confidence, we empirically demonstrate the growth of ionized regions
during the reionization epoch for the first time. We independently estimate the
upper limit on the Str\"omgren radii (\Rstrom), and derive the escape fraction
of ionizing photons (\fesclyc) from the ratio of \Rbub/\Rstrom, deriving a
median value of 5\% which on average represents the lower end of the photon
budget necessary for reionization. 
\end{abstract}


\section{Introduction}

Observations of \lya\ emission from galaxies have long been known to fulfill a
key role in charting the history of cosmic reionizaton \citep[see][for a
review]{Dijkstra.2014}.  Because nebular \lya\ emission can be redshifted from
the systemic velocity (typically by a few hundred \kms) it can be detected
through the damping wing of the Gunn-Peterson trough
\citep{Miralda-Escude.Rees.1998}, albeit with possibly-significant attenuation.
This makes the population \lya-emitting galaxies a very powerful tool to study
the epoch of reionization (EoR), where an abundant \lya-emitting galaxy
population produces strong line emission just redward of systemic velocity.

Purely photometric \lya\ measurements have been employed for EoR studies for
almost two decades.  The most common approaches have been to study flux deficit
of \lya\ compared with expectations. The first attempts performed differential
comparison of the \lya\ luminosity function (LF) across redshifts
\citep[e.g.][]{Malhotra.2004,Kashikawa.2006}, although see
\citet{Dijkstra.2007lf} for further considerations. Developments of the
technique study \lya\ in comparison to the UV continuum flux, either as the the
`volumetric escape fraction' \citep{Hayes.2011evol,Wold.2017} or an evolution
of the equivalent width distribution
\citep{Stark.2011,Schenker.2014,Cassata.2015,Arrabal-Haro.2018}.  These
methods, however, all rely upon ensemble of galaxies to derive one quantity,
which is typically the average neutral fraction (\xHI) at a given redshift
\citep[e.g.][]{Mason.2018igmfrac}. One cannot trivially derive higher order
estimates of the reionization process, such as spatial variations, bubble size
distribution, etc.  For this kinematic/spectroscopic data are needed. 

More nuanced estimates can be attained if the intrinsic \lya\ properties of a
galaxy are known: specifically comparisons of the observed \lya\ EW and
velocity profiles with their intrinsic values would lead directly to a measure
of the distance between a galaxy and the foreground screen of absorbing \hI.
Little progress has been made because we need to know both the EW and velocity
offset of emergent \lya, which depend on stellar conditions and radiative
transfer effects in the interstellar and circumgalactic media
\citep{Verhamme.2006,Dijkstra.2006,Laursen.2009}.  Moreover, without systemic
redshifts (\zsys) we cannot even begin to estimate the velocity offset with
respect to that of the IGM (simply distance in an expanding universe).
\citet{MasonGronke.2020} provide the singular exception to this, attempting to
circumvent the problem using a double-peaked \lya-emitter and deriving a bubble
radius of $\approx 0.7$~pMpc in the $z=6.6$ galaxy COLA1 \citep{Matthee.2018}.

However this field is rapidly changing, and over recent years systemic
redshifts have become available from other far ultraviolet emission lines like
\cIII]~$\lambda\lambda 1907,1909$\AA\
\citep{Stark.2015,Stark.2017,Mainali.2017} and infrared lines like
[\cII]158\micron\ \citep{Pentericci.2016,Carniani.2017,Endsley.2022rebels}.
More recently, JWST has delivered the restframe optical emission lines to
provide \zsys\ and also find \lya\ emission at $z>7.2$
\citep{Tang.2023,Saxena.2023} and even almost $z= 11$ \citep{Bunker.2023}.
Thus, part of the requirement of having systemic redshifts and \lya\ EWs and
velocity offsets are now falling into place. 

In this Letter we take advantage of the fact that \lya\ EWs and velocity
offsets are now being measured in the EoR. We study a sample of 23 galaxies at
$z=6-11$ with systemic redshifts and \lya\ EWs and velocity shifts, which we
present in Section~\ref{sect:sample}.  Their emergent \lya\ profiles (i.e.
those that leave the galaxy after radiation transport in the ISM and CGM) are
not known, but using a large sample of low-$z$ galaxies \citep{Hayes.2023lya}
where IGM attenuation is negligible, we build realistic models for the \lya\
that emerges from galaxies. While there is no guarantee that the emergent \lya\
spectral profiles at low-$z$ match those in the EoR, we showed in
\citet{Hayes.2021} that we do not find evidence of their evolution in
currently-available data.  Using the expected damping wing from a neutral
universe, we build a model for the emergent \lya\ observables, fitting the size
of the ionized region and emergent \lya\ EW in a hierarchical Bayesian
framework.  Thus, we empirically derive the distribution of the sizes of
ionized bubbles that surround galaxies across most of the reionization
timeline.  This method is described in Section~\ref{sect:methods} and the
results in Section~\ref{sect:results}.  We discuss the impact of various
assumptions in Section~\ref{sect:uncertainties}, and present our concluding
remarks in Section~\ref{sect:conclusions}.  Throughout we assume a cosmology of
$\{H_0,\Omega_\mathrm{M},\Omega_\Lambda\} =
\{70~\mathrm{km~s}^{-1}~\mathrm{Mpc}^{-1},0.3, 0.7\}$.

\section{The Galaxy Sample} \label{sect:sample}

\begin{figure}
\noindent
\begin{center}
\includegraphics[width=0.99\linewidth]{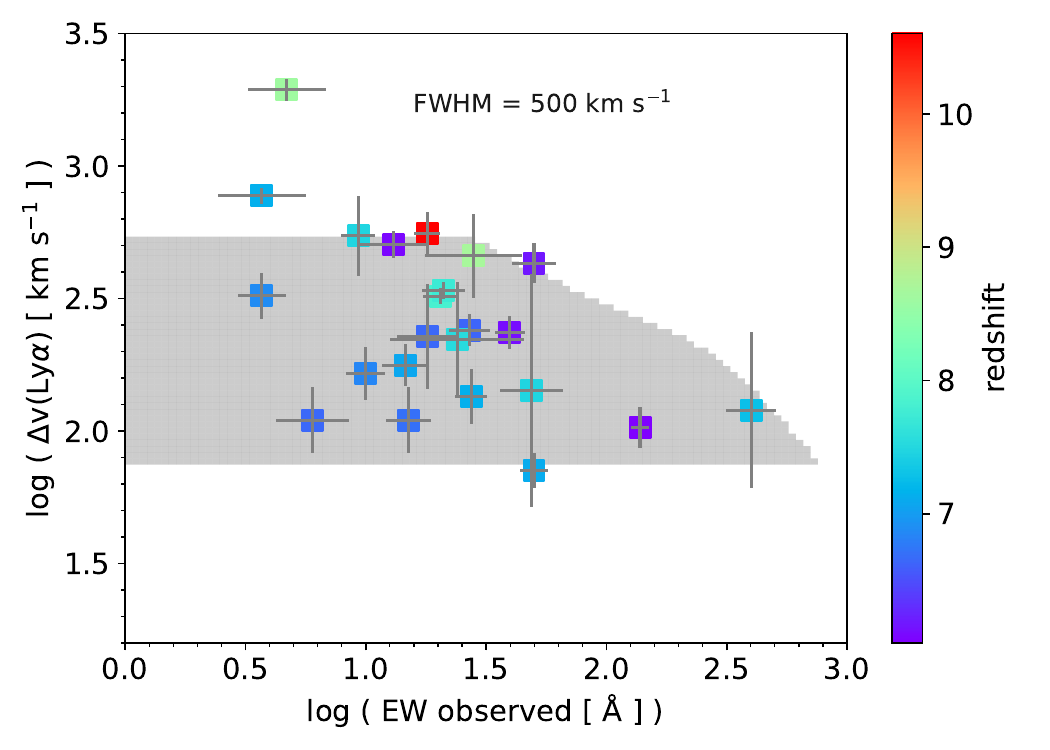}
\end{center}
\caption{The \lya\ observables of our sample. \dvred\ is plotted on the
ordinate axis, with EW on the abscissa. Galaxy redshift is color-coded, with
the colorbar to the right.  The shaded region shows the parameter space that
can be matched using the emergent line profile of a low-$z$ galaxy
\citep{Hayes.2023lya}, when broadened synthetically to FWHM~$=500$~\kms.  We
assume the IGM density at $z=7$, vary the velocity from 0 to 3000~\kms, and the
EW between 0 and the very high value of 800~\AA. }
\label{fig:mom1_vs_ew}
\end{figure}

We obtain the spectral measurements for 23 galaxies at $z>6$, by combining
various literature samples. We include the fifteen galaxies compiled by
\citet{Endsley.2022rebels}, searching the primary references to obtain
uncertainties on the \lya\ EW and its velocity offset, \dvred.  We add eight
more galaxies with recently-obtained measurements from JWST.  Six of these are
taken from the The Cosmic Evolution Early Release Science Survey
\citep[CEERS,][]{Finkelstein.2023} that have \lya\ detections: four from
\citet{Tang.2023} at $z=7.4-8.7$, and two from \citet{Jung.2023} at $z\simeq
7.5$ after excluding one system for which \zsys\ is discussed as uncertain.
The final two targets are JADES targets GNz11 at $z\simeq 10.6$
\citep{Bunker.2023} and GS-z7-LA at $z\simeq 7.3$ \citep{Saxena.2023}.  The
main properties of interest are \lya\ EW, velocity offset of the red peak,
systemic redshifts and UV magnitudes.  We show the distribution of these
properties in Figure~\ref{fig:mom1_vs_ew}.

We note that this is a compilation of sources reported in different surveys,
and requires spectroscopic detections in both \lya\ and non-resonant emission
lines. The quantitative interpretation will naturally be prone to selection
effects. For example, over such a broad redshift range  Malmquist biases are
possible, but within this small sample there is currently no systematic
evolution in the average luminosity. Different emission lines also measure
\zsys\ at different redshifts, potentially impacting redshift precision where
only weak lines are visible, but these uncertainties are treated within our
method (Section~\ref{sect:modeling}).  There is a general trend for galaxies
with larger velocity shifts to show smaller EWs: this trend has been noted
before, and is demonstrated in low-$z$ samples \citep[e.g.][]{Hayes.2023lya}
where the IGM has no influence.  The relation probably arises because more
massive galaxies have higher column densities of neutral gas: \lya\ photons
must therefore undergo more scattering events in order to take longer frequency
excursions to the wings of the line profile and see the gas as optically thin
\citep[e.g.][]{Verhamme.2008,Hashimoto.2013}.  This results in larger velocity
offsets and also smaller \lya\ escape fractions, because of the increased
probability of dust absorption.

As the IGM becomes thicker with increasing redshift, and ionized regions are
presumably smaller, a trend of increasing \dvred\ with redshift could be
expected.  However, the apparent trend for galaxies with larger velocity shifts
to lie at $z\gtrsim 7.5$ is not found to be significant by a two-sample
Kolmogorov-Smirnoff test.

\begin{figure*}
\noindent
\begin{center}
\includegraphics[width=0.54\linewidth]{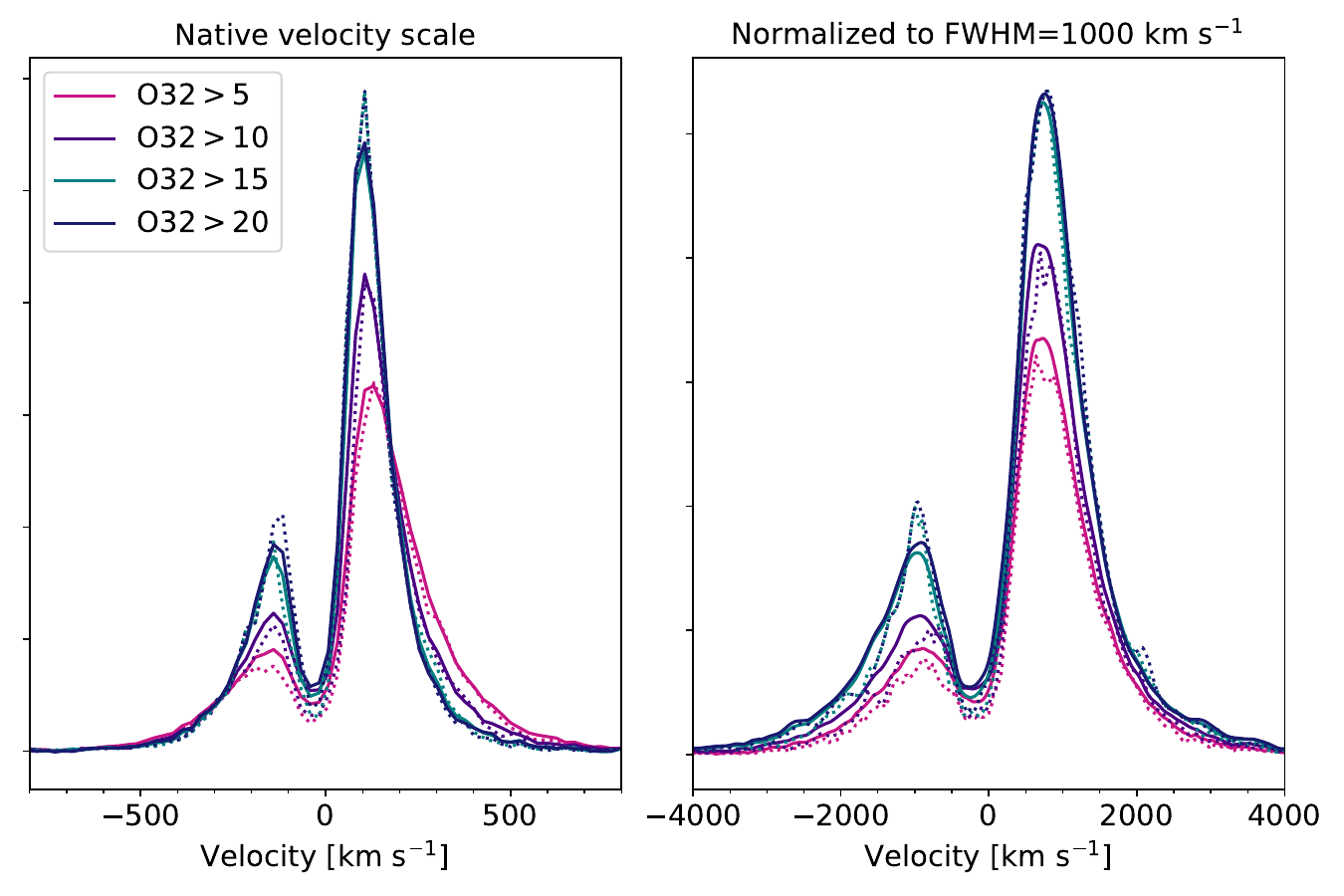}
\includegraphics[width=0.44\linewidth]{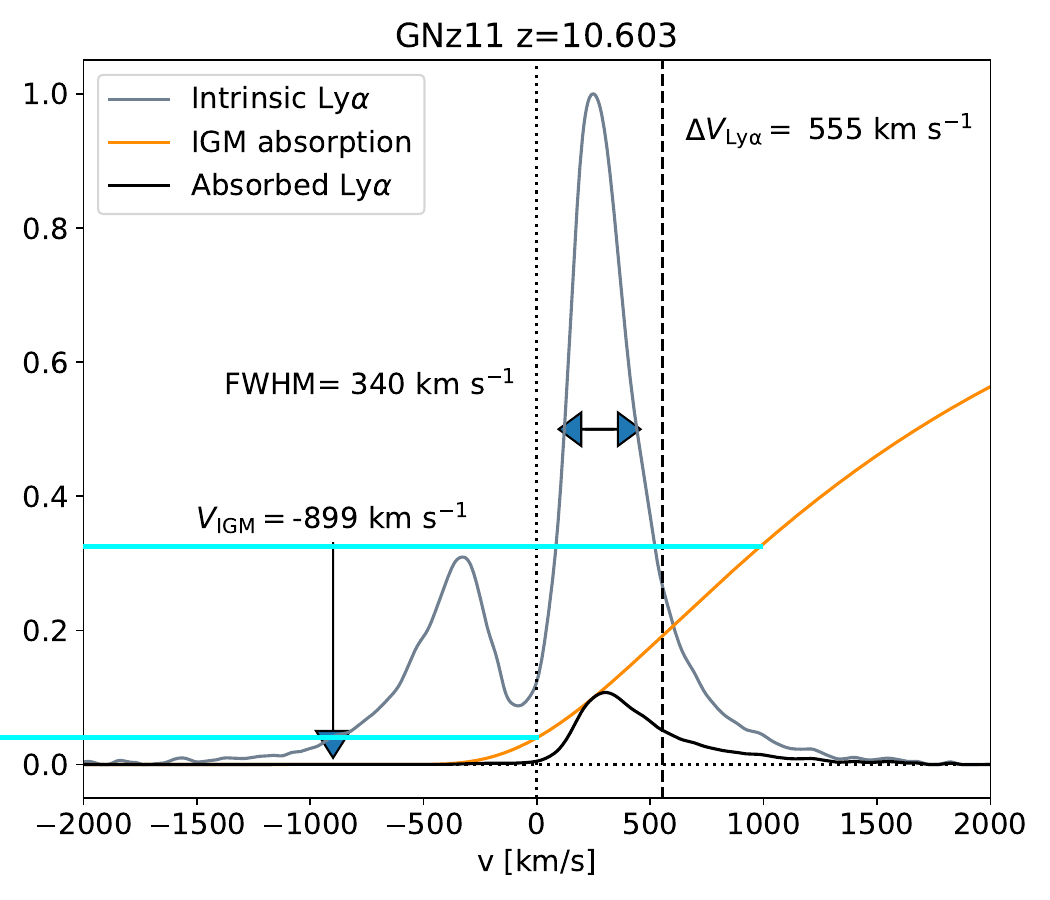}
\end{center}
\caption{Left: model \lya\ profiles.  The far left spectrum shows the stacked
profiles, where each spectrum is simply renormalized in luminosity at $\Delta
v>0$ before stacking.  As illustration, each colored line shows the stack of a
subsample where all galaxies exceed the labeled O$_{32}$ ratio: greater than 5
in pink to greater than 20 in black.  The central spectrum shows the same as
the left, but the spectra are also rescaled to the same FWHM of the red peak
prior to stacking.  The reference FWHM is set to 1000~\kms\ (note the different
abscissa range of the two plots).  Solid lines are mean stacks and dotted lines
are median. The far right panel shows an example profile for GNz11 before and
after IGM attenuation. We show the emergent \lya\ profile as the grey line and
the shape of the best-recovered Gunn Peterson absorption in orange. The
observed \lya\ profile is the product of the two and is shown in black.
Various characteristic velocities are labeled.}
\label{fig:lyaprofiles}
\end{figure*}

\section{Modeling Galaxies at Redshift 6--11} \label{sect:methods}

\subsection{The `Emergent' Lyman alpha profile}\label{sect:intrinsic_profile}

We use three different adjectives to describe \lya\ in this paper: the
\emph{intrinsic} properties that are produced by the \hII\ regions, and the
\emph{observed} properties that reach the telescope are commonly-used terms.
Here we also use the term \emph{emergent}, which refers to the \lya\ emitted by
the galaxy (after the CGM) but before attenuation by the IGM.  The emergent
\lya\ profile is estimated from the sample of starburst galaxies at
$0.05<z<0.44$ observed with the Cosmic Origins Spectrograph on HST
\citep{Hayes.2021,Hayes.2023lya}.  At this redshift, the \lya\ profile is
unaffected by IGM absorption. 

In \citet{Hayes.2023lya} we produced stacked average \lya\ spectra, where we
binned the sample by various galaxy properties.  Here we use the same software
to estimate the emergent \lya\ profiles from the EoR galaxies, and show some
examples in the left panels of Figure~\ref{fig:lyaprofiles}.  We stack spectra
based up measured quantities that match galaxies recently observed with
JWST/NIRspec \cite[e.g.][]{Brinchmann.2022,Cameron.2023,Tang.2023}, such as
high [\oIII]$\lambda 5007$/[\oII]$\lambda 3727$ line ratios ($\equiv
\mathrm{O}_{32}$).  We adopt the stack of galaxies with $\mathrm{O}_{32}$
ratios above 10, to approximately match the measured values at $z>6$ where
these measurements have been made \citep{Tang.2023,Saxena.2023}.  While the
optical spectroscopic properties of the low- and high-$z$ galaxies match well,
the study relies upon invariance of the \lya\ profiles with redshift, which we
cannot test directly for EoR galaxies.  However we showed in \citet{Hayes.2021}
that these profiles accurately reproduce those of \lya-emitters at $3 \lesssim
z \lesssim 6$ when the effects of an absorbing IGM are applied.  We regard this
as cause for optimism, and assert that the same general profile shape can be
applied at higher redshifts still.

All spectra are first continuum-subtracted, using the modeled continuum spectra
described in \citet{Hayes.2023lya} and \citet{Hayes.2023wind}.  We then shift
all spectra into the restframe, using the systemic redshifts measured from
optical emission lines -- average stacks of these spectra are shown in the left
panel of Figure~\ref{fig:lyaprofiles}, where each has been normalized by the
luminosity in the red \lya\ peak before combination.  However, because we are
mostly interested in the spectral shape of the \lya\ profile at velocities
redwards of line-centre, we also re-normalize our spectra onto a common
frequency metric: we convert the spectra to velocity space, and rescale each to
FWHM=1000~\kms\ in the red peak -- these are shown in the central panel of
Figure~\ref{fig:lyaprofiles}.  The individual resampled spectra are renomalized
by redshifted \lya\ luminosity before combination.  This provides us with the
average shape of the \lya\ emission, and in computing it we have written up a
function to give the \lya\ profile for an arbitrary input FWHM.  Thus, our
software  treats FWHM as a free parameter.

\subsection{Absorption by the Intergalactic Medium}\label{sect:igm_model}

With a model for the emergent \lya\ profiles, we attenuate the spectra with a
model IGM.  We take the cosmic hydrogen number density scaled by a factor of
$(1+z_\mathrm{sys})^3$ at each redshift.  From this we calculate the expected
\lya\ optical depth profile $\tau_\mathrm{IGM}(v)$ as a function of velocity,
as radiation redshifts through a neutral universe.  

The unknown quantity is by how much the \lya\ is cosmologically redshifted
before it encounters the absorbing \hI\ in the foreground.  We estimate the
absorption of \lya\ using Voigt profiles, implementing successive absorptions
numerically over velocity shift, \vigm, which equates to a distance in an
expanding universe.  We multiply the emergent \lya\ profile by $\exp
(-\tau_\mathrm{IGM}(v))$ and calculate the zeroth and first moments of the
`observed' profile.  For a given emergent EW we rescale the zeroth moment to an
`observed' EW that accounts for the IGM.  I.e $EW_\mathrm{obs} =
EW_\mathrm{emerge}  (m_\mathrm{0,obs}/m_\mathrm{0,emerge})$.
$EW_\mathrm{emerge}$ therefore becomes a free parameter in our model, and can
be compared with the EWs and velocity shifts shown in
Figure~\ref{fig:mom1_vs_ew}.

\begin{figure*}
\noindent
\begin{center}
\includegraphics[width=0.99\linewidth]{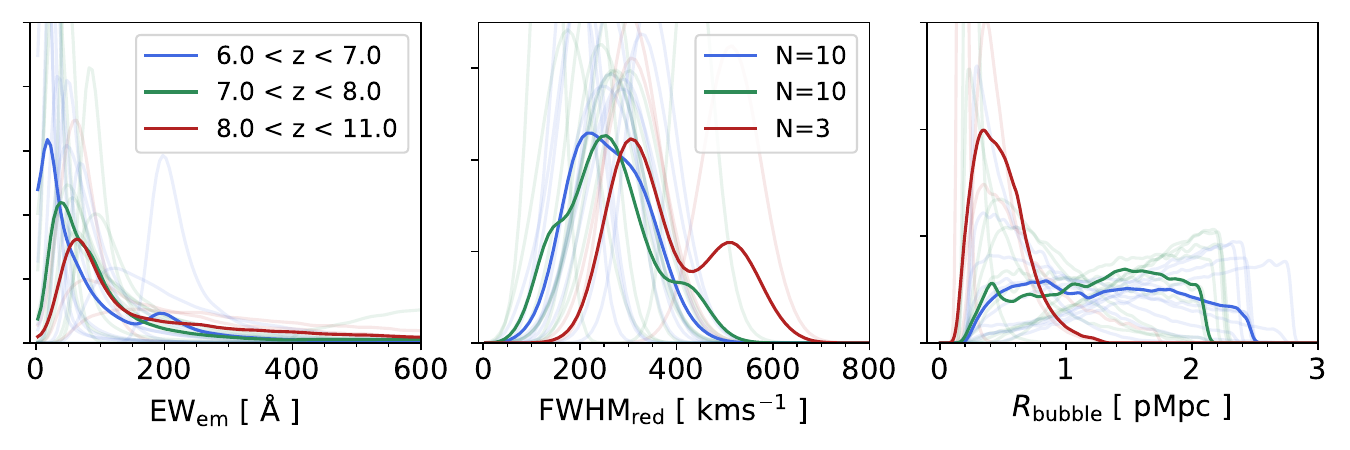}
\end{center}
\caption{Binned probability distribution functions from traces of the Monte
Carlo sampler. Left shows the emergent \lya\ EW, and center shows the FWHM of
the \lya\ red peak.  Right shows the bubble size in proper Mpc.  Each posterior
is color-coded by redshift, as labeled in the left plot.  Hard lines show the
means of the posterior distributions in each redshift bin, while faint lines
show those for individual galaxies.  The number of galaxies per bin is shown in
the legend of the central plot.}
\label{fig:pdfs}
\end{figure*}

\subsection{Bayesian Inference on EW$_{\rm em}$, FWHM$_{\rm red}$, and \Rbub}\label{sect:modeling}

For each galaxy, we perform a hierarchical Bayesian inference analysis to
estimate the free parameters of our model:  $EW_{\rm em}$, $FWHM_{\rm red}$,
and \Rbub. We model the likelihood of observing the available data, i.e., the
pair of $\{ EW_\mathrm{obs}\pm
\sigma_{EW_\mathrm{obs}},\Delta(v_\mathrm{Ly\alpha})_\mathrm{obs}\pm\sigma_{\Delta(v_\mathrm{Ly\alpha})}
\}$, with a bivariate Normal distribution including known errors. Defining the
vector of model parameters as $\theta=\{EW_{\rm em}, FWHM_{\rm red},v_{\rm
IGM}\}$, we  write the posterior for $\theta$ as: \noindent \begin{multline*}
\hspace{-5mm} p(\theta|
EW_\mathrm{obs},\sigma_{EW_\mathrm{obs}},\Delta(v_\mathrm{Ly\alpha})_\mathrm{obs},\sigma_{\Delta(v_\mathrm{Ly\alpha})},
M_\mathrm{UV})\propto \\ \hspace{-5mm}
N(EW_\mathrm{tr},\Delta(v_\mathrm{Ly\alpha})_\mathrm{tr}|\sigma_{EW_\mathrm{obs}},\sigma_{\Delta(v_\mathrm{Ly\alpha})},\theta,M_\mathrm{UV})p(\theta|M_\mathrm{UV})
\label{ew_distribution} \end{multline*}

\noindent
where $p(\theta |M_{UV})$ is the prior on the model parameters, given the UV
magnitude of each galaxy. Subscript `tr' refers to \lya\ quantities after
transmission through the IGM, which are compared to the observed quantities
given a subscript `obs'.  We assume that  $p(\theta
|M_\mathrm{UV})=p(EW_\mathrm{em})p(v_{\rm IGM})p(FWHM_{\rm
red}|M_\mathrm{UV})$, i.e., the priors on the emergent EW, the \lya\ line width
and the bubble radius are independent.  The prior on \vigm\ is the simplest to
treat -- we have no empirical  knowledge of this and adopt a uniform prior. For
the emergent FWHM distribution we base our prior upon strong trends between
$M_\mathrm{UV}$ and the FWHM of the red peak \citep{Hayes.2023lya}.  We adopt
measurements from the \emph{Lyman alpha Spectral Database}
\citep[LASD;][]{Runnholm.2021}\footnote{\href{http://lasd.lyman-alpha.com}{http://lasd.lyman-alpha.com}}
and the GALEX-measured UV magnitudes \citep[see][Figure 21 of the online-only
material]{Hayes.2023lya} and fit a power-law between FWHM and $M_\mathrm{UV}$;
our prior is then Normal around this relation, and includes errors on the fit.
For the emergent \lya\ EW we adopt an exponential distribution based upon very
deep MUSE and HST observations at $z=6$.  We take the exponential scale length
of $EW_0=212\pm 186$\AA\ from  \citet{Hashimoto.2017}. This value is already
corrected for IGM absorption in that paper (following \citealt{Inoue.2014},
same as \citealt{Hayes.2021}), and thus should be comparable to the emergent EW
required for our method.  We then make the assumption that the EW distribution
does not evolve strongly over the redshift of the sample.  

We sample the posterior distribution using a Metropolis Hastings sampler built
in \textsc{pycm} \citep{pymc}.  \vigm\ is converted to \Rbub\ using Hubble's
law.  We show an example fit in the right panel of
Figure~\ref{fig:lyaprofiles}.  In Figure~\ref{fig:pdfs} we show the posterior
distribution functions for $EW_{\rm em}$, $FWHM_{\rm red}$, and \Rbub\ for each
galaxy (light curves), divided into three samples according to redshift.  In
the discussion that follows, we use the maximum a posteriori and its associated
68\% credibility interval as our best estimate and associated uncertainty on
each parameter.

\section{Main Results and Discussion} \label{sect:results}

\subsection{An Example: GNz11}

We show an example model for GNz11 \citep{Bunker.2023} in the right panel of
Figure~\ref{fig:lyaprofiles}.  This is currently the highest redshift known
\lya-emitting galaxy. The galaxy has an observed \lya\ EW of 18~\AA, and
\dvred\ of 555~\kms.  While it may be an AGN, our models show that its spectral
properties in \lya\ can be recovered with relatively normal conditions.
Naively, it may be considered difficult to explain a \lya\ emission line from
$z= 10.6$, but in actuality it is not especially hard.  

The maximum of the posterior on the \lya\ FWHM is 340~\kms\ -- this is quite
high, but within the range of widths measuerd in low-$z$ galaxies (the broadest
we find in the $z\sim 0$ COS sample has a FWHM=350~\kms\ in its red peak).
This modeled broad \lya\ has a wing visible out to $v \simeq 1200$~\kms\ at low
flux density.  The maximum of the posterior probability function for the IGM
velocity  is almost 900~\kms\ -- at this velocity offset, the damping wing
draws a steep diagonal line across the emerging red \lya\ peak, absorbing $\sim
8$ times more flux at $v=0$ compared to at $v=1000$~\kms.  The result is a
significant shift in the first moment of the line, to the 555~\kms\ that is
observed.  In doing so the \lya\ is also significantly suppressed, and only
$\sim 7$\% of the emergent \lya\ survives IGM absorption.  The emergent \lya\
EW is $\simeq 260$~\AA\ with a credibility range of 153--800~\AA\ at 16--84\%.
This value is again high for a star-forming galaxy, but not higher than
observed in galaxies at effectively all redshifts.  Based upon the observed
\hgamma\ flux and Case B recombination theory, \citet{Bunker.2023} estimate
that around 4\% of the \emph{intrinsic} \lya\ escapes the galaxy.  The is
comparable to the 7~\% we estimate from the \emph{emergent} \lya\ flux; the
further factor of $\sim 2$ could indicate that $\sim 50$~\% of the intrinsic
\lya\ escapes the galaxy after internal radiative transfer effects.

\subsection{Sample-Averaged Galaxy Properties}

We take the \hI\ velocity offsets, emergent \lya\ EW and FWHM  from the maximum
values of the posterior sampling (Section~\ref{sect:modeling}) and show the
pdfs in Figure~\ref{fig:pdfs}. Faint lines show the pdfs for individual
galaxies, color coded by redshift, while solid lines show the arithmetic mean
pdf in each bin.  Emergent EWs are typically peaked towards the lower end of
the distribution near 100~\AA\ -- a value that is quite typical of star-forming
galaxies and does not require extreme stellar populations or AGN.  Only a
handful of individual galaxies show posteriors that peak at EWs above 200~\AA,
one of which is obviously JADES-GS-z7-LA, with its observed EW of 400~\AA.

The FWHM also peak in ranges that are typical of star forming galaxies at
low-$z$, of around 200~\kms, with a tail up to 600~\kms.  Here there is a more
obvious signature for galaxies at the higher redshifts to be intrinsically
broader than those at $z=6-8.5$, although we note that this highest redshift
bin contains only three galaxies.  This may be a signature of a selection bias
in the requirement of \lya, where only broad emergent \lya\ profiles are able
evade the damping wing of the IGM and be reported as a \lya\ detection.

\subsection{The Evolution of Ionized Regions Through the EoR}\label{sect:sizeevol}

After converting the IGM velocity offset to a foreground \hI\ distance using
the Hubble parameter, we show the resulting combined posterior probabilities in
the right panel of Figure~\ref{fig:pdfs}.  Estimated \hI\ foreground distances
generally show flat combined posteriors for the $z=6-8$ subsamples, which
indicates a broad range of bubble sizes, from $\simeq 0.5$~pMpc, and extending
to $\simeq 2.5$~Mpc.  However the combined posterior of the $z>8$ subsample is
much more strongly peaked around distances below 1~pMpc, shows a much sharper
decline with redshift, and drops to zero at 1~pMpc. 

\begin{figure}
\noindent
\begin{center}
\includegraphics[width=0.99\linewidth]{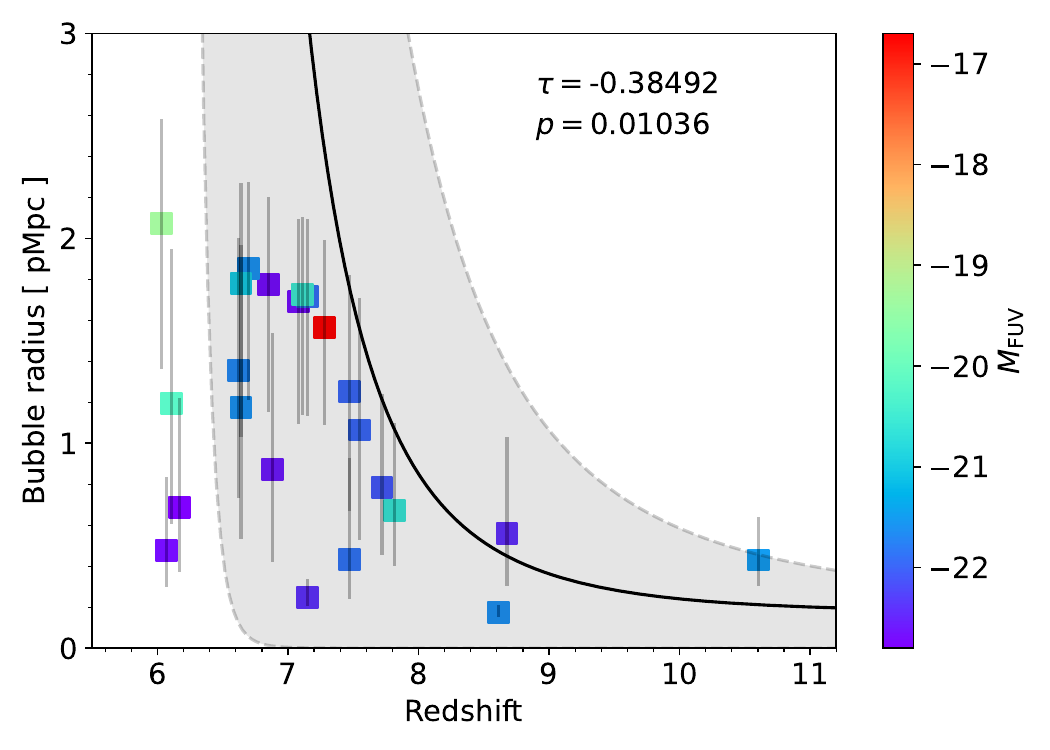}
\end{center}
\caption{Evolution of the bubble radius, \Rbub\ (ordinate) with redshift
(abscissa).  Uncertainties refer to the 16th and 84th percentiles of the
posterior.  Absolute UV magnitude is color-coded.  The shaded regions shows the
10-90-percentile range of bubble sizes from the simulation of
\citet{Giri.2021}. }
\label{fig:Rbub_vs_z}
\end{figure}

\begin{figure*}
\noindent
\begin{center}
\includegraphics[width=0.49\linewidth]{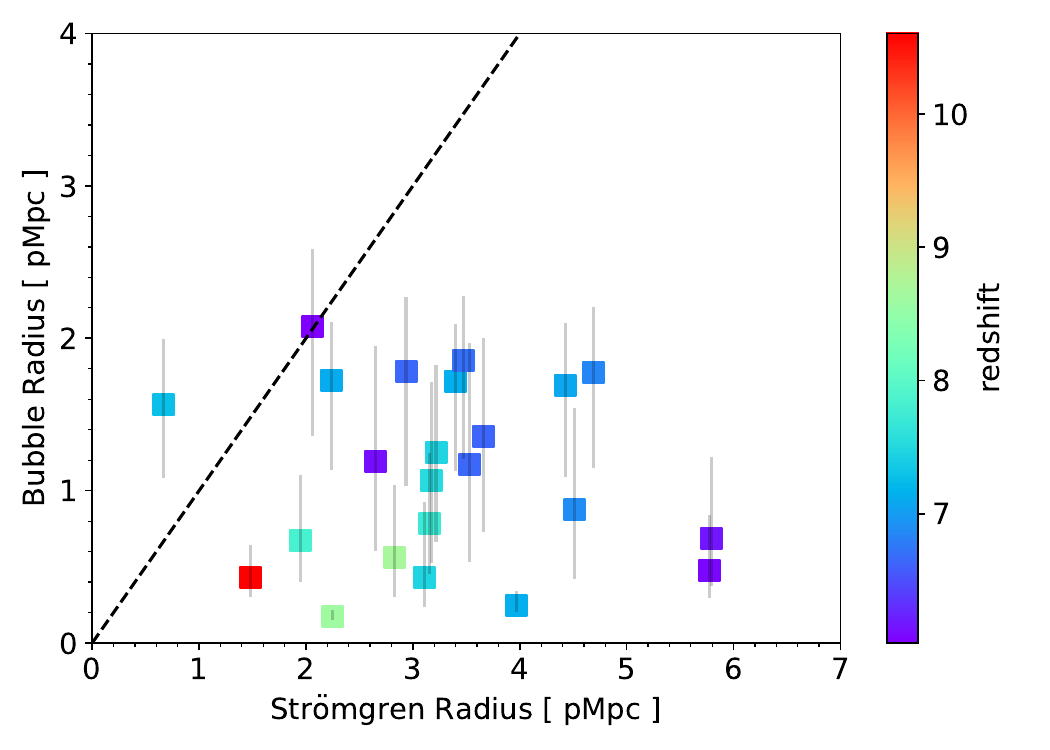}
\includegraphics[width=0.49\linewidth]{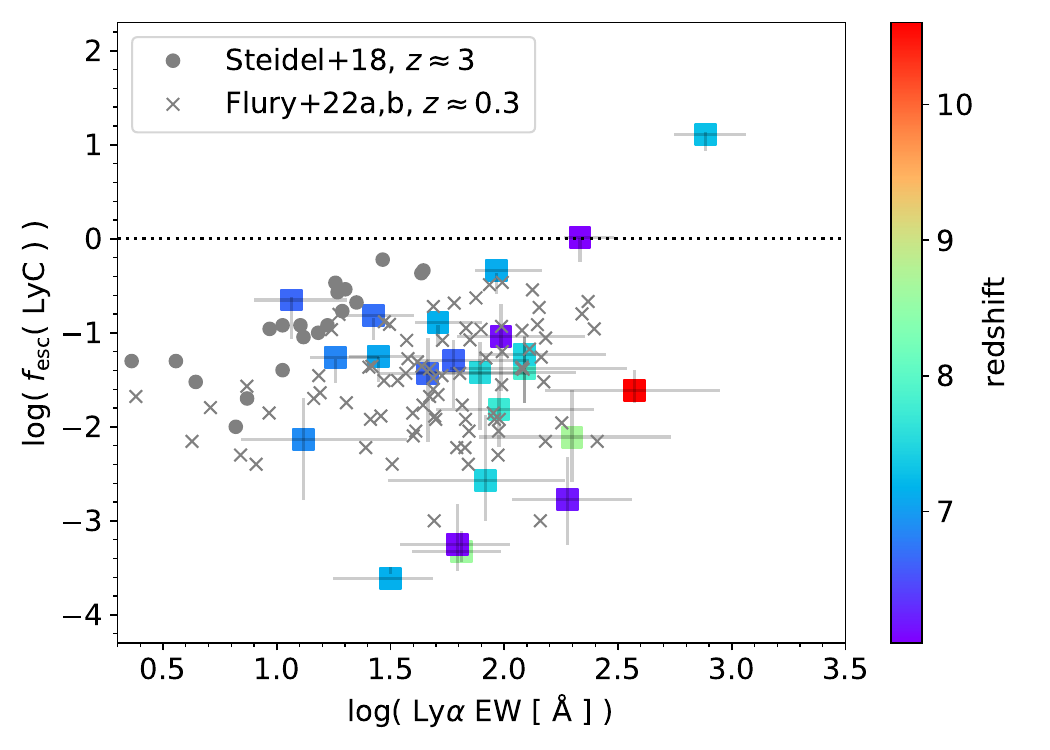}
\end{center}
\caption{Left: the radii of the bubble inferred from \lya\ modeling (\Rbub),
contrasted with that of the Str\"omgren sphere (\Rstrom).  Right: the
comparison of the LyC escape fraction (\fesclyc; inferred from the left plot)
with the model-inferred EW of \lya.  We over-plot the values measured at
$z\simeq 3$ by \citet{Steidel.2018} and at $z\simeq 0.3$ by
\citet{Flury.2022pap1}.  Redshift is color-coded in both cases.}
\label{fig:fesc}
\end{figure*} 

In Figure~\ref{fig:Rbub_vs_z} we show the evolution of the distance to the
foreground screen of \hI\ (taking the peak of the posterior) as a function of
redshift.  Distances to the foreground \hI\ screen range from around 0.5 to 2.5
proper Mpc at the lower redshift end.  We note that our approach -- and
observations in general -- may not be sensitive to smaller bubble radii, as a
certain minimum offset will be required for \lya\ to be detected, and therefore
be reported in the literature.  There is in principle no reason why the method
could not be applied to galaxies without \lya\ emission, but in order to avoid
great degeneracies such an approach would require more informed priors on the
intrinsic EW.  This could perhaps be attained from optical line emission
\cite[e.g.][]{Runnholm.2020,Hayes.2023lya}.

The interpretation of the larger radii at $z\simeq 6$ is  likely to be that the
these galaxies reside in an ionized universe, and far out in the damping wing
(after $\sim 2000$~\kms) there is little power in the IGM to modulate the \lya\
observables -- the method breaks down for large bubbles and all \vigm\ are
equally valid.  The lower end of the distribution (\vigm~$\simeq 1$~pMpc) are
consistent estimates from \lya\ spectral profiles \citep{MasonGronke.2020} and
from ionization balance considerations \citep{Bagley.2017}, both of which are
estimated at comparable redshifts.

Figure~\ref{fig:Rbub_vs_z} is striking in its absence of larger bubbles at
higher redshifts.  Using Kendall's $\tau$ statistic, we find the trend to be
significant at $p =1 \times 10^{-2}$, and we note that similar credibility
regions are determined at $z>8$ compared with the galaxies at the lower
redshift end of the sample.  If taken at face value, the result implies that
ionized regions increase in size from $z\simeq 11$ to 6 -- this empirical find
of ionized regions growing with time fulfills the main expectations of
reioinization.  The selection bias of needing \lya\ in emission may also enter
here: smaller bubbles will not enable \lya\ transfer through the IGM, implying
a lower limit on the bubble sizes we can capture. As this is more likely to be
the case at earlier times, the strength of the relationship would only
increase.

In the same figure we also show computational results from \citet{Giri.2021}.
The simulated relation takes a similar form: starting from $z\sim 11$, the
bubble size of GNz11 falls just at the upper edge of the distribution.  The
simulated bubble sizes then increases towards $z\simeq 7$, forming a fully
ionized universe (effectively infinite bubble radii) by $z=6.5$.  However,
simulated bubbles grow somewhat faster than our data suggest.  There are four
observed galaxies at $z\sim 6$ that seemingly have much smaller bubble radii
than suggested by the simulations.  If the universe is fully ionized at $z=6$
on these sightlines, the inference for these galaxies could be the result of
proximate \lya-absorbing systems (possibly galaxies) whose damping wings
suppress and redshift \lya, and cause our model estimate smaller radii.  This
result may reflect the high end of the broader range of \lya\ optical depths
observed in quasar spectra at $z\approx 6$ \citep[e.g.][]{Bosman.2022}. 

\subsection{The Escape of Ionizing Radiation}

In Section~\ref{sect:sizeevol} we estimated the distance between the galaxy and
the absorbing \hI\ gas in the foreground.  Here we test whether these main
targeted galaxies are capable of ionizing their own \hII\ regions and, if so,
what their properties must be. We proceed by simply investigating the size of
the Str\"omgren sphere: \Rstrom~$=(3 Q_0 f_\mathrm{esc}^\mathrm{LyC} / 4\pi
n_\mathrm{H}^2 \alpha_\mathrm{B})^{1/3}$.  Here, $Q_0$ is the intrinsic
production rate of ionizing photons and \fesclyc\ is the ionizing escape
fraction; hence the product of the two is the emitted ionizing photon rate.
$\alpha_\mathrm{B}$ is the (temperature dependent) total recombination rate
coefficient under Case B, for which we assume $10^4$~K. This includes three
main assumptions: (1.) that we can estimate $Q_0$ from data; (2.) that galaxies
remain ionizing for long enough to ionize their surrounding media; (3.) the IGM
is homogeneous and of fixed density.

We assume \nh\ to be the cosmic value at the redshift of each galaxy, as we
described also for the inference in Section~\ref{sect:igm_model}. Each galaxy
must be strongly star-forming, as evidenced by the emergent \lya\ EWs of
100~\AA\ (intrinsic values are likely higher still).  Addressing assumption 1
above, we assume the ionizing photon production efficiency (\xiion) measured
for GNz11 of $5.2 \times 10^{25}$~Hz~erg$^{-1}$ \citep{Bunker.2023} and convert
the UV luminosity to $Q_0$.  By first setting \fesclyc\ to 1, we compute
$R_\mathrm{S}^ \mathrm{max}$, finding values between 0.5 and 6 Mpc. We contrast
this with the distances to the foreground \hI\ (0.5--2.5~pMpc) in the left
panel of Figure~\ref{fig:fesc}.  It is immediately obvious that, if the above
assumptions hold, then 22 of the 23 galaxies have sufficient ionizing power to
ionize their own bubble.

Obviously \fesclyc\ cannot be 1, or the galaxies would not produce the strong
\lya\ or observed nebular line emission. \fesclyc\ cannot be measured at these
redshifts, but under the assumption that each galaxy is singularly responsible
for ionizing its own \hII\ region, we can solve for \fesclyc\ in each galaxy by
setting \Rstrom\ equal to the inferred bubble size:
$f_\mathrm{esc}^\mathrm{LyC} = (R_\mathrm{Bub}/R_\mathrm{S})^3$.  We show
\fesclyc\ against the inferred \lya\ EW in the right panel of
Figure~\ref{fig:fesc}.  These two quantities are strongly correlated at low-
and mid-redshifts and we overplot the individual data points of
\citet{Flury.2022pap2} and stacks and \citet{Steidel.2018} for comparison.  Our
estimates span the same range as the lower redshift estimates.  While this plot
is suggestive that relationship could also be identifiable for galaxies in the
EoR, we do not yet have sufficient samples to make similar statements. The
median value of \fesclyc\ in our sample is 5.1~\%, with a 16--84 percentile
range of 1--21\%, which is broadly consistent with requirements for cosmic
reionization \citep[e.g.][]{Finkelstein.2019}.  For GNz11 we estimate \fesclyc\
to be just 2.4~\% with a $1\sigma$ confidence intervals between 1 and 6.1\%.
This is remarkably coincident with the value of $0.03^{+0.05}_{-0.02}$,
estimated by \citet{Bunker.2023} using completely different methods. One galaxy
significantly outlies the distribution, with \Rbub\ almost 3 times its supposed
Str\"omgren radius, leading to an apparent escape fraction of 13.  This is the
extreme galaxy GS-z7-LA \citep{Saxena.2023}, with \lya\ EW of 400\AA, and
demonstrating that our assumptions do not hold everywhere.

\section{Sources of Uncertainty}\label{sect:uncertainties}

We now discuss each of the assumptions above as a source of uncertainty,
beginning with the ionizing photon production efficiency. We have assumed
log(\xiion/Hz~erg$^{-1}$)~$=25.7$ throughout.  In our formulation
\fesclyc~$\propto \xi_\mathrm{ion}^{-1}$, and variations of a factor of 2 in
\xiion\ will change \fesclyc\ by the corresponding amount.  Significant
reductions in \xiion\ are hard to envisage, as they must be able to reproduce
the strong emergent \lya\ and nebular line emission observed in the available
JWST spectra. Increasing \xiion\ by a factor of 2 may perhaps be possible
\citep[e.g.][]{Maseda.2020}, but values higher than this would become
inconsistent with predictions from normal stellar populations.

Secondly we have assumed the Str\"omgren sphere can fully form, which would
require $\simeq0.6-7$~Myr based upon the inferred range of \Rbub, and is
largest at the lower redshift end. This timescale is short compared to the
typical ionizing timescales of galaxies, even at this epoch. By modeling the
SEDs of galaxies at a similar epoch, \citep{Endsley.2022nircam} find typical
stellar ages greater than this for $\simeq80$~\% of their sample (and also
exactly the same median value of \xiion\ we adopt above). Moreover, starburst
events should not coordinate on times shorter than the freefall timescale,
$t_\mathrm{ff}$: to bring $t_\mathrm{ff}$ below 7~Myr for a mass of
$10^8$~\msun\ would require all gas to fall from a radius of just 250~pc, which
we deem unrealistic.

Next we address the assumption of fixed \nh: \fesclyc\ is proportional to the
square of this density. It is worth noting that recombination timescale of gas
at cosmic \nh\ is about twice the Hubble time at $z=6$: global recombinations
do not significantly affect our calculations. If \nh\ varies substantially, it
will also impact our estimates of \Rbub, and to test this we have re-run our
inference with \nh\ rescaled from the cosmic average. Reducing \nh\ by a factor
of 2 decreases the median \Rbub\ by only 16\%, but increases \Rstrom\ by 60\%;
consequently \fesclyc\ decreases to a median value of $\simeq1$\%. However it
is very unlikely that the vicinity of galaxies is underdense at all compared to
the cosmic average. The reverse situation is more likely: doubling \nh\
marginally increases \Rbub, while \Rstrom\ decreases, causing the median
\fesclyc\ to increase to $\simeq 18$\%, which remains a realistic value in the
EoR.  In the instance where gas is clumped on smaller scales within the bubble,
dense regions will experience increased recombination rates while the regions
between them will have lower densities, which allows ionization fronts to
propagate faster.  In this case, sightline effects could also become important:
if the denser regions lie in front of a galaxy, the excess absorption will push
the inferred \Rbub\ to larger values to compensate. Assuming the denser clumps
occupy a small volume, the decreased \nh\ elsewhere will mean the true \Rstrom\
is slightly larger than our estimate.  In this instance \fesclyc\ will be
underestimated, although if the denser regions lie outside of our sightline,
the effects will mostly cancel.

Finally, we address the question of multiple galaxies contributing to the
ionization of a single \hII\ region, which is indeed likely.  However, it is
also probable that the UV-selection of these targets has found the most
luminous galaxy in the vicinity. Our median value of \Rbub~$=1.1$ physical Mpc
corresponds to a volume of $\simeq 2850$ comoving Mpc$^3$ at $z=7$, and such a
volume would on average contain only 0.04 galaxies brighter than \Lstar\
according to the recent LF of \citet{Harikane.2023}.  In an unclustered
universe, we would have to integrate to luminosities $\sim 5$ times fainter for
the probability of finding another galaxy within \Rbub\ to reach 1. With all
other quantities held constant, this would add only 20\% more ionizing
luminosity and not substantially changing the results.  Of course galaxies do
cluster, but we expect these results to hold on average in cases where the most
luminous local source has been identified.  We expect, however, that this is
not the case for the very faint galaxy GS-z7-LA \citep{Saxena.2023}, which has
an observed \lya\ EW of $\simeq400$\AA, and almost certainly requires
assistance from nearby ionizing sources.

\section{Concluding Remarks} \label{sect:conclusions}

We have built a hierarchical Bayesian model to estimate the intrinsic \lya\
observables (emergent equivalent width and kinematic offset from systemic
velocity) of galaxies in the reionization epoch, as well as the size of the
ionized regions (`bubbles') in which they must reside.  The model is built upon
empirical \lya\ spectral templates observed in lower redshift galaxies where
the IGM has no impact, and estimates the IGM absorption at a given redshift
that best matches observation.  We have applied this framework to a sample of
23 ostensibly star-forming galaxies at redshift $z=6-11$ where systemic
redshifts are available, including very recent observations from JWST.  We find
the following main results. 

\begin{itemize}
\item{The observed galaxies occupy ionized regions with sizes between 0.5 and
2.5 proper Mpc.  The posterior probability distribution of the bubble size is
not invariant with redshift, and is skewed towards smaller bubbles at higher
redshifts.  We detect an upwards evolution of the bubble size with redshift
that is significant at better than $3\sigma$ and demonstrates that ionized
regions grow with time.  The recovered bubble radii are broadly consistent with
results from numerical simulations of reionization.}
\item{From the observed UV luminosity and reported ionizing photon production
efficiencies, we compute the size of the Str\"omgren radius of each galaxy.
The Str\"omgren radius does not correlate with the bubble radius.  We use the
ratio of these two radii to estimate the escape fraction of ionizing photons,
recovering a median value of 5\% -- this is marginally consistent with the
requirements for galaxies to reionize the universe.  If the IGM density within
the ionized region is overdense by a factor of 2, \fesclyc\ increases to 18\%.
Our estimates of \fesclyc\ and \lya\ EW are comparable to those derived at
lower-$z$, but we do not yet recover the correlation between these quantities.}
\end{itemize}

\section*{Acknowledgements}
In non-specific, groupwise order we thank (1.) Sambit Giri, Ivelin Georgiev and
Garrelt Mellema for making simulation data available for
Figure~\ref{fig:Rbub_vs_z}; (2.) Axel Runnholm and Max Gronke for useful
discussions and further ongoing contributions to the projects;  (3.) The
anonymous referee whose careful reading of the manuscript has improved the
rigor of this work.  M.H. acknowledges the support of the Swedish Research
Council, Vetenskapsr{\aa}det and is Fellow of the Knut and Alice Wallenberg
Foundation.  

\bibliographystyle{aasjournal}

\clearpage

\end{document}